\documentclass[namedreferences]{solarphysics}
\usepackage[hyperref,optionalrh,solaromanenum]{spr-sola-addons} 
\usepackage{graphicx}                    
\usepackage{amssymb}                    
\usepackage{color}                       
\usepackage{breakurl}                    
\begin{document}


\begin{opening}

\title{}

%
\author[addressref={aff1},corref,email={jojuta@gmail.com}]{\inits{J.J.}\fnm{Jouni}~\lnm{Takalo}}

\institute{$^{1}$ Space Physics and Astronomy Research Unit, University of Oulu, POB 3000, FIN-90014, Oulu, Finland
}
%
\runningauthor{J.J. Takalo}
\runningtitle{Distribution of solar corona for Cycles 18-24}



\begin{abstract}
Homogeneous coronal data set (HCDS) of the green corona (Fe XIV) and coronal index of the solar activity (CI) have been used to study time-latitudinal distribution in solar cycles 18\,--\,24 and compared with similar distribution of sunspots, the magnetic fields and the solar radio flux 10.7 cm. The most important results are: (a) distribution of coronal intensities related to the cycle maximum are different for individual cycles, (b) the poleward migration of the HCDS from mid latitudes in each cycle exists, even in extremely weak Cycle 24, and the same is valid for the equatorward migration (c) the overall values of HCDS are slightly stronger for the northern hemisphere than for the southern one, (d) distribution of the HDCS are in coincidence with strongest photospheric magnetic fields (B$>$50 Gauss) and histogram of the sunspot groups, (e) Gnevyshev gap was confirmed with at least 95\,\% confidence in the CI, however, with different behavior for odd and even cycles.
Principal component analysis (PCA) showed that the first and second component account for 87.7\,\% and 7.3\,\% of the total variation of the CI. Furthermore, the PC2 of the green corona was quite different for cycle 21, compared with other cycles.
\end{abstract}

%

\end{opening}


\section{Introduction}

Green 530.3 nm (Fe XIV) line is the most prominent coronal irradiance indicator. Its importance is due to existence all over the solar limb and the interconnection to the local strength of the magnetic fields of the Sun \citep{Rusin_2002}. This line originating at a temperature of about 2 million K. Its first casual observations were made soon after the invention of a coronagraph in 1939 at the Arosa observatory by \cite{Waldmeier_1957}; more systematic observations started in 1946 at several coronal stations worldwide \citep{Minarovjech_2011_b}. Quite continuous measurements of green corona exist thus for Solar Cycles 18\,--\,23. In addition, there exists now data for Solar Cycle 24 \citep{Lukac_2010}.

\cite{Kane_2015} deals with the similarities and differences of the coronal index of solar activity (CI) and sunspot numbers (SSN) using normalized indices. He states that both have very similar shape in the ascending and maximum phase of the solar cycle, but differ in the descending phase of the cycle. This is due to varying abundance of coronal holes in the descending phase of the cycles. The Cycles 19 and 22 are most identical in shape with the CI and SSN parameters. Cycles 18 and 23 have excess CI compared to SSN and Cycles 20 and 21 have excess CI but located at the end of the descending phase. This shows that the shape of the CI is not related to 22-year Hale cycle.

Another guestion is, does corona exhibit so-called Gnevyshev gap (GG), which has been reported in many solar phenomana \citep{Gnevyshev_1977, Feminella_1997, Ahluwalia_2004, Bazilevskaya_2006, Kane_2008a, Norton_2010, Du_2015, Takalo_2018, Takalo_2020_a} and even in the geomagnetic indices \citep{Takalo_2021_b}. This has not been reported earlier. \cite{Rybansky_2001}, however, stated that the existence of double-maxima as found by \cite{Gnevyshev_1967} was not confirmed in the corona index (CI). Single maximal peaks of the CI were mostly observed in coincidence with sunspot numbers, even though some time shift could occur with a comparison of Sunspot number, e.g. of 2 years in cycle 21.

This article is organized as follows: Section 2 presents the data and methods used in this article. In Section 3 we study spatial and temporal statistics using the homogeneous coronal data set (HCDS), i.e. 72 values with 5 degree resolution around the solar limb starting from north pole and circulating counterclockwise (see more in Data and Methods section). In Section 4 we explore the coronal index of solar activity (CI) in order to find clues about GG in the corona, and compare sunspot numbers and solar radio flux to CI using Principal component analysis (PCA). We discuss the results and give our conclusion in Section 5.

\section{Data and methods}

\subsection{Corona indices}

The homogeneous coronal data set (HCDS) is the irradiance of the Sun as a star in the
coronal green line (Fe XIV, 530.3 nm). It is derived from ground-based observations of the green corona made by the network of coronal stations (Kislovodsk, Lomnick\'{y} \v{S}t\'{i}t, Norikura, and Sacramento Peak). These indices are not, however, measured anymore in the traditional way as was made earlier at Lomnick\'{y} \v{S}t\'{i}t Observatory (former Lomnick\'{y} \v{S}t\'{i}t coronal station). The coronal intensities have been measured at 72 points at 5 degree separation starting from north pole counterclockwise around the Sun at height around 50 arcsec. The values are calibrated to the center of the solar disk to get absolute values of intensity, i.e. absolute coronal units (ACU). One ACU represents the intensity of the continuous spectrum of the center of the solar disk in the width of one {\AA}ngstr{\"o}m at the same wavelength as the observed coronal spectral line (1ACU = 3.89 Wm$^{-2}$ sr$^{-1}$ at 530.3 nm). We call this data later sometimes as 'latitudinal corona index'.

\cite{Rybansky_1975} introduced the coronal index of solar activity (CI) as a general index
of solar activity. CI, a full-disk index, represents the averaged daily irradiance emitted through the green coronal line into one steradian towards the Earth. It is expressed in power units Wsr$^{-1}$ as measured from ground-based observatory. The idea is that these values can be transferred to other units in order to compare them to satellite measurements. The monthly values of CI are between 2\,--\,20 $\times$ 10$^{16}$ Wsr$^{-1}$, and daily values have, thus far, been always under 30 \citep{Rybansky_1975, Rybansky_2001, Rybansky_2005, Minarovjech_2011_a}.

Recently the corona indices were corrected mainly for pre-1966 era. In this research we use the new reconstructed corona indices, which were fetched from https://www.ngdc.noaa.gov/stp/solar/corona.html \citep{Rybansky_2005}. The main period in this study is Solar Cyles 18\,--\,23, in order to have same amount of even and odd cycles. In some cases we also use data for Solar Cycle 24. These are fetched from  http://www.suh.sk/obs/vysl/MCI.htm \citep{Lukac_2010}.


\subsection{Sunspot Groups}

When plotting sunspot groups we use the data set of sunspot groups for Solar Cycles 18\,--\,24 by \cite{Leussu_2017}. This data set contains latitude and time stamp of sunspot groups as seen for the first time. This data set does not include whole Solar Cycle 24, but only years to the end of 2016. This is why the period 2017-2019 is missing. The minima and length of the sunspot cycles used in this study for solar and also geomagnetic indices are listed in Table 1.

\begin{table}
\small
\caption{Sunspot-Cycle lengths and dates [fractional years, and year and month] of (starting) sunspot minima for Solar Cycles 12\,--\,23 (except the end of data for Solar Cycle 24). \citep{NGDC_2013}.}
\begin{tabular}{ c  c  l  c }
  Sunspot cycle    &Fractional    &Year and month     &Cycle length  \\
      number    &year of minimum   & of minimum     &    [years] \\
        \hline   
18    & 1944.1  &1944 February  & 10.2  \\
19    & 1954.3  &1954 April  & 10.5  \\
20    & 1964.8  &1964 October  & 11.7  \\
21    & 1976.5  &1976 June & 10.2  \\
22    & 1986.7  &1986 September  & 10.1  \\
23    & 1996.8  &1996 October  & 12.2  \\
24    & 2009.0  &2008 December  & 10.9  \\ 
25    & 2020  &2019 December 
\end{tabular}
\end{table}

\subsection{Two-Sample T-Test}

The two-sample T-test for equal mean values is defined as follows. The null hypothesis assumes that the means of the samples are equal, i.e. $\mu_{1}=\mu_{2}$. Alternative hypothesis is that $\mu_{1}\neq\mu_{2}$. The test statistic is calculated as
\begin{equation}
T = \frac{\mu_{1} - \mu_{2}}{\sqrt{{s^{2}_{1}}/N_{1} + {s^{2}_{2}}/N_{2}}} ,
\end{equation}
where $N_{1}$ and $N_{2}$ are the sample sizes, $\mu_{1}$ and $\mu_{1}$ are the sample means, and $s^{2}_{1}$ and $s^{2}_{2}$ are the sample variances. If the sample variances are assumed equal, the formula reduces to
\begin{equation}
T = \frac{\mu_{1} - \mu_{2}} {s_{p}\sqrt{1/N_{1} + 1/N_{2}}} ,
\end{equation}
where
\begin{equation}
s_{p}^{2} = \frac{(N_{1}-1){s^{2}_{1}} + (N_{2}-1){s^{2}_{2}}} {N_{1} + N_{2} - 2} .
\end{equation}
The rejection limit for two-sided T-test is $\left|T\right| > t_{1-\alpha/2,\nu}$, where $\alpha$ denotes significance level and $\nu$ degrees of freedom. The values of $t_{1-\alpha/2,\nu}$ are published in T-distribution tables \citep{Snedecor_1989, Krishnamoorthy_2006, Derrick_2016}.  Now, if the value of p$<\alpha=0.05$, the significance is at least 95\%, and if p$<\alpha=0.01$, the significance is at least 99\%.

\subsection{Principal component analysis method}

Principal component analysis is a useful tool in many fields of science including chemometrics \citep{Bro_2014}, data compression \citep{Kumar_2008} and information extraction \citep{Hannachi_2007}. PCA finds combinations of variables, that describe major trends in the data. PCA has earlier been applied, e.g., to studies of the geomagnetic field \citep{Bhattacharyya_2015}, geomagnetic activity \citep{Holappa_2014_2, Takalo_2021_b}, ionosphere \citep{Lin_2012}, the solar background magnetic field \citep{Zharkova_2015}, variability of the daily cosmic-ray count rates \citep{Okpala_2014}, and atmospheric correction to cosmic-ray detectors \cite{Savic_2019}.

In this article we compare Sunspot number (SSN) and Solar 10.7 cm Radio Flux data (SRF) for Solar Cycles 19\,--\,24 to the same cycles of coronal index of solar activity (CI). To this end, we estimate that the average length of the cycle is 130 months, and use it as a representative Solar Cycle. We first resample the monthly data such that all cycles have the same length of 130 time steps (months), i.e about the average length of the Solar Cycles 19\,--\,24 \citep{Takalo_2018, Takalo_2020_a, Takalo_2021_b}. This effectively elongates or abridges the cycles to the same length. Before applying the PCA method to the resampled cycles we standardize each individual cycle to have zero mean and unit standard deviation. This guarantees that all cycles will have the same weight in the study of their common shape. Then after applying the PCA method to these resampled and standardized cycles, we revert the cycle lengths and amplitudes to their original values.

As said earlier, we standardize each individual cycle to have zero mean and unit standard deviation. Standardized data are then collected into the columns of the matrix $X$, which can be decomposed as \citep{Hannachi_2007, Holappa_2014_1, Takalo_2018}

\begin{equation}
	X = U\:D\;V^{T}  \     ,
\end{equation}

where $U$ and $V$ are orthogonal matrices, $V^{T}$ a transpose of matrix $V$, and $D$ a diagonal matrix 
	$D= diag\left(\lambda_{1},\lambda_{2},...,\lambda_{n}\right)$
with $\lambda_{i}$ the $i^{th}$ singular value of matrix $X$. The principal component are obtained as the the column vectors of

\begin{equation}
P  = U\!D.
\end{equation}
	
The column vectors of the matrix $V$ are called empirical orthogonal functions (EOF) and they represent the weights of each principal component in the decomposition of the original normalized data of each cycle $X_{i}$, which can be approximated as

\begin{equation}
	X_{i} = \sum^{N}_{j=1} \:P_{ij}\:V_{ij} \   ,
\end{equation}

where j denotes the $j^{th}$ principal component (PC). The explained variance of each PC is proportional to square of the corresponding singular value
$\lambda_{i}$. Hence the $i^{th}$
PC explains a percentage
\begin{equation}
\frac{\lambda^{2}_{i}}{\sum^{n}_{k=1}\!\lambda^{2}_{k}} \cdot\:100\%
\end{equation}
of the variance in the data.

\section{Homogeneous Coronal Data Set}

\begin{figure}
	\centering
	\includegraphics[width=1.0\textwidth]{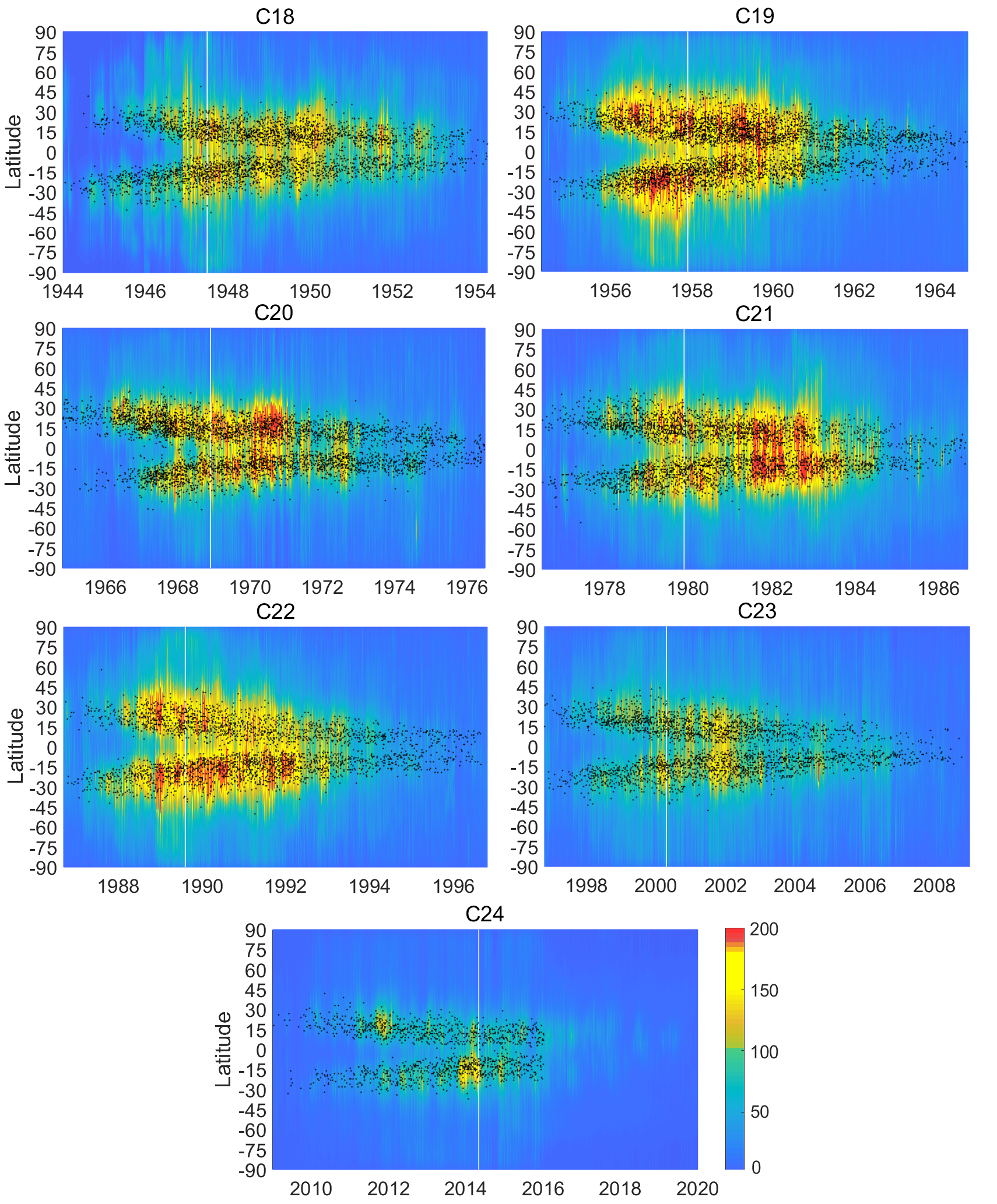}
		\caption{A combined figure of homogeneous corona data set (color bar) and sunspot groups (black dots) separately for Solar Cycles 18\,--\,24. The white vertical lines show the maxima of the cycles. (Note that sunspots for 2017-2019 are missing from the used database.)}
		\label{fig:Sunspots_and_corona_18_24}
\end{figure}

Figure \ref{fig:Sunspots_and_corona_18_24} shows the sunspot groups and homogeneous coronal data set (HCDS) indices \citep{Minarovjech_2011_b} separately for the Solar Cycles 18\,-\,24.  The intensity of these coronas are shown as a color bar on the right side of Cycle 24 panel. It seems that the most intense corona is around the maximum for Cycles 19 and 22, but are spread wider for the other cycles. Notice also that the largest sunspots are also concentrated around the maximum for the Cycles 19 and 22, but are more abundant in the descending phase for other cycles, especially so for the Cycle 21. The maxima of the cycles are depicted with vertical white line. The used database contains incomplete Cycle 24 and has groups only to the end of the year 2016 \citep{Leussu_2017}.

\begin{figure}
	\centering
	\includegraphics[width=1.0\textwidth]{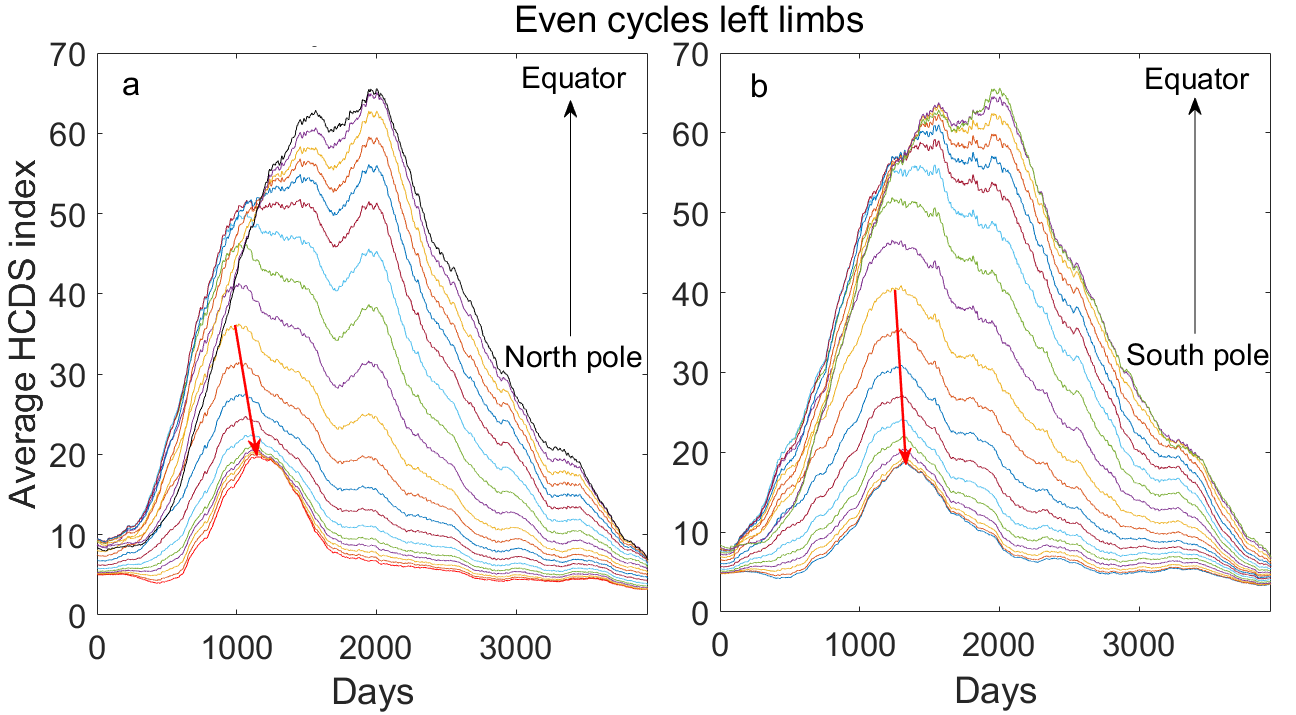}
		\caption{The lines (with 5-degree resolution) of the average HCDS indices in the left limb of the Sun as a function of day for the average even cycle. The lines are arranged such that the lowest lines are at poles and highest lines at equator. The red arrows show the temporal drift of the latitudinal maxima towards poles. Figs. \ref{fig:HCDS_lines}a and b are for northern and southern hemisphere, respectively. }
		\label{fig:HCDS_lines}
\end{figure}

\begin{figure}
	\centering
	\includegraphics[width=1.0\textwidth]{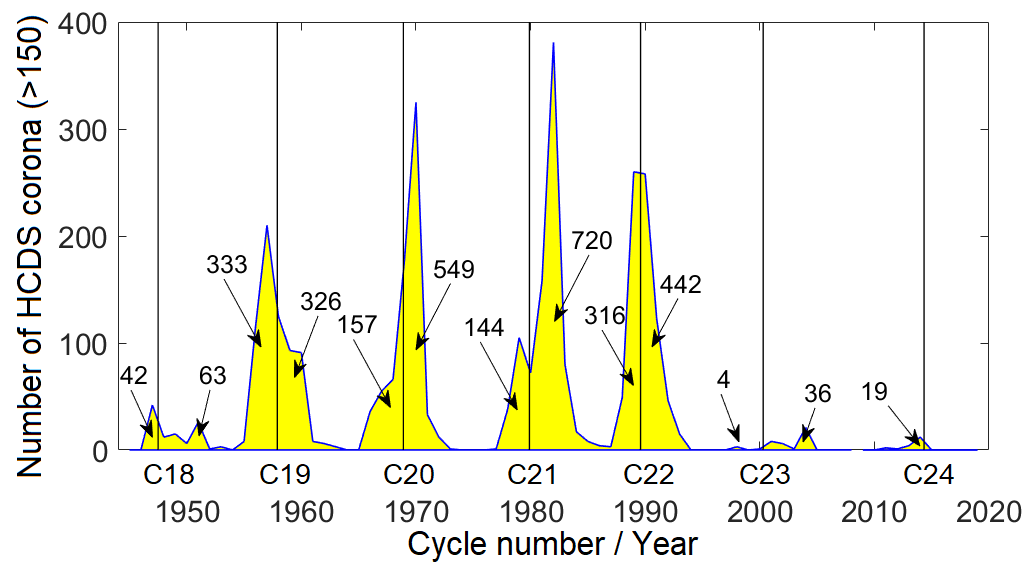}
		\caption{Number of HCDS indices greater than 150 for the Cycles 18\,--\,24. The black vertical lines show the maxima of the cycles.}
		\label{fig:lat_corona_150}
\end{figure}

Figure \ref{fig:HCDS_lines}a and b the lines of average HCDS indices of the left limbs of the Sun as a function of day of the average even cycle for northern and southern hemisphere of the Sun, respectively. The lines with 5-degree separation are arranged such that the lowest lines are at poles and highest lines near equator of the Sun. The red arrows show the drift of the latitudinal maximum towards the poles. The maximum HCDS corona maximum appears first around 1000 days, i.e. about a quarter of the cycles near 40-50 degrees of northern latitude and somewhat later in the southern hemisphere at corresponding latitudes. The maxima appear at the pole around 1200 days and 1350 days after the start of the cycle for northern and southern hemisphere, respectively \citep{Minarovjech_2011_a}. Similar drift but stronger appears towards equator in both hemispheres. These reach their maxima at about 2000 days, i.e. at the halfway of the cycle . Notice, however, that there exists another, local maximum at about 1560 days after the start of the cycle, and a gap between these two maxima. We believe that this is related to the Gnevyshev gap. For odd cycles the migration of the latitudinal maximum is similar, although somewhat more complex. This is probably due to huge difference in the distribution of the HCDS corona for odd cycles. For example Cycle 21 has two maxima at the north pole first at 1350 days and another late in the descending phase at about 2500 after the start of the cycle. 
Figure \ref{fig:lat_corona_150} confirms that cycles 19 and 22 are quite symmetric around the maximum in the number of strong HCDS corona ($>$150) events. Note also that Cycle 24 has only 19 strong corona events and all of those are before the maximum in April 2014.

\begin{figure}
	\centering
	\includegraphics[width=1.0\textwidth]{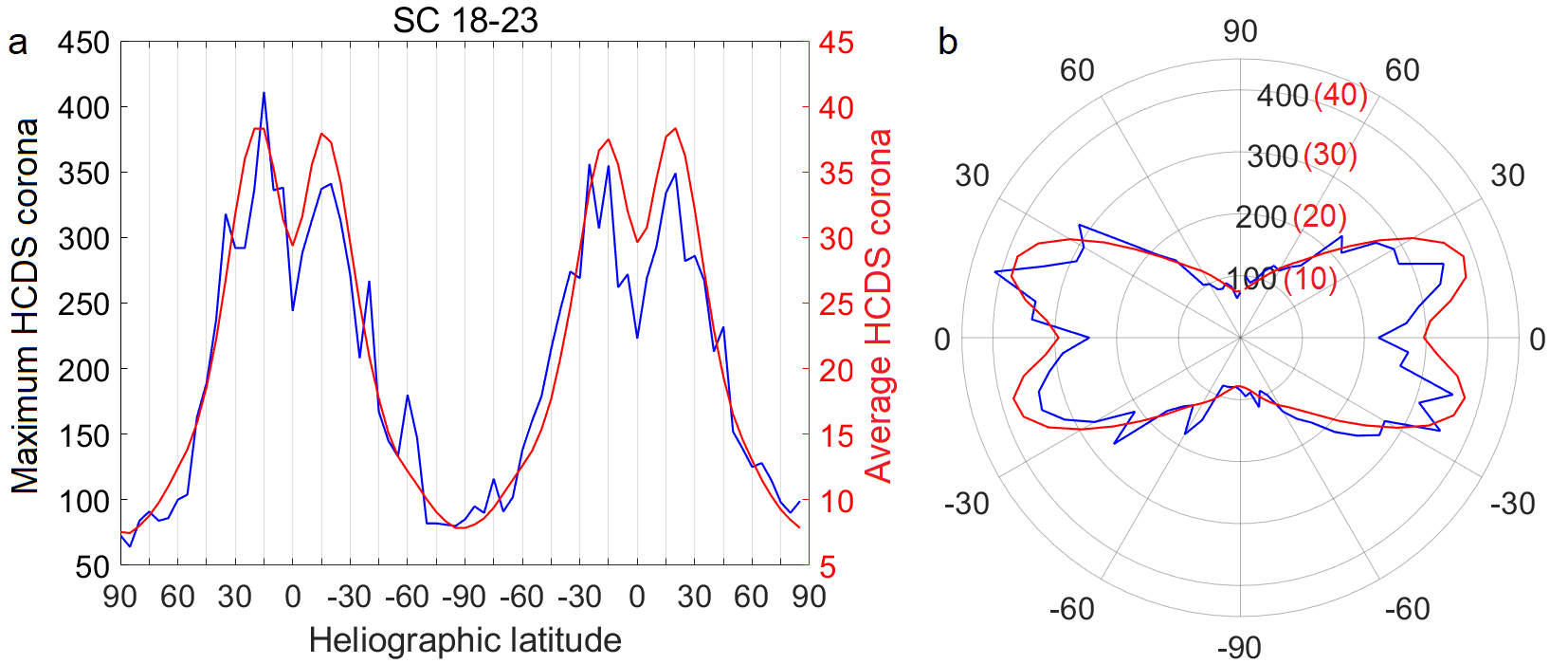}
		\caption{The maximum (blue, left vertical axis) and average values (red, right vertical axis) of HCDS for Solar Cycles 18\,--\,23 a) in Cartesian coordinates, and b) in polar coordinates. (Note that for polar coordinates maximum is black value and average red value.)}
		\label{fig:maximum_and_avg_corona}
\end{figure}

\begin{figure}
	\centering
	\includegraphics[width=1.0\textwidth]{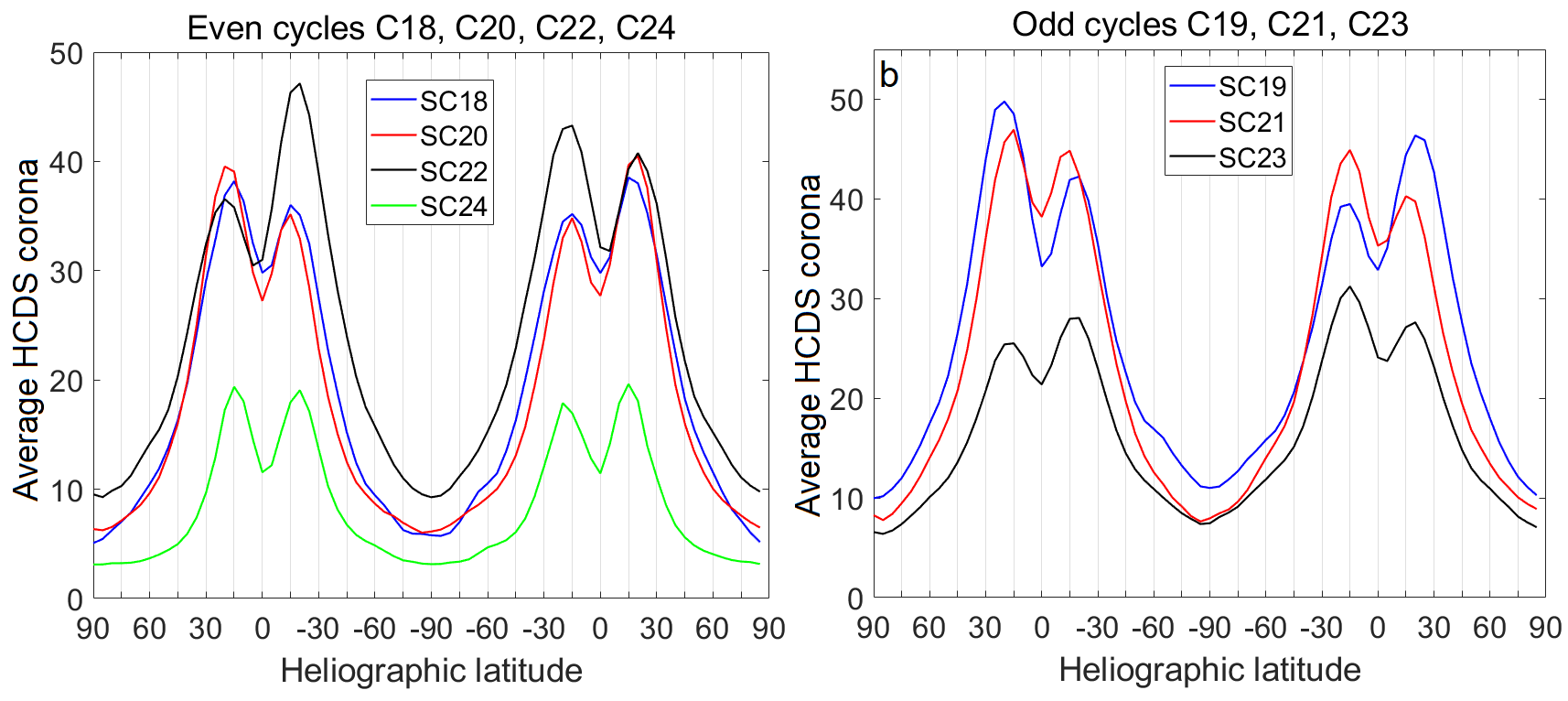}
		\caption{The average values of latitudinal (HCDS) corona for Solar Cycles 18\,--\,24 a) for even cycles, and b) for odd cycles.}
		\label{fig:avg_corona}
\end{figure}

Figure \ref{fig:maximum_and_avg_corona} shows the maximum (blue) and average (red) values of HCDS corona for Solar Cycles 18\,--\,23 in Cartesian (a) and polar (b) coordinates. The left and right sides of the figure (here and in the succeeding figures) show the values at left and right limb of the sun, respectively. It is evident that the absolute values of the maxima are all between 15\,--\,20 degrees of heliographic latitude. These latitudes are somewhat larger than the maxima of the sunspot distributions, which are between 14.9 and 15.5 for even and odd cycles, respectively \citep{Takalo_2020_b}. The overall values of the corona are slightly stronger for northern hemisphere than the southern hemisphere. The issue of northern/southern hemisphere asymmetry is, however, more complicated than this simple figure displays, (see more, e.g., \cite{Dzifcakova_1998, Joshi_2015}).
Figure \ref{fig:avg_corona} shows the latitudinal distribution of the HCDS corona for even and odd cycles separately. It is noticeable that the Cycles 19 and 22 are most asymmetrical such that northern hemisphere is dominating for Cycle 19 and southern hemisphere is dominating for Cycle 22. These are also the cycles, which were mentioned earlier as the most symmetric around the maximum for strong corona events. The corona values for Solar Cycles 23 and 24 are understandably by far smaller than for the other cycles.

\begin{figure}
	\centering
	\includegraphics[width=1.0\textwidth]{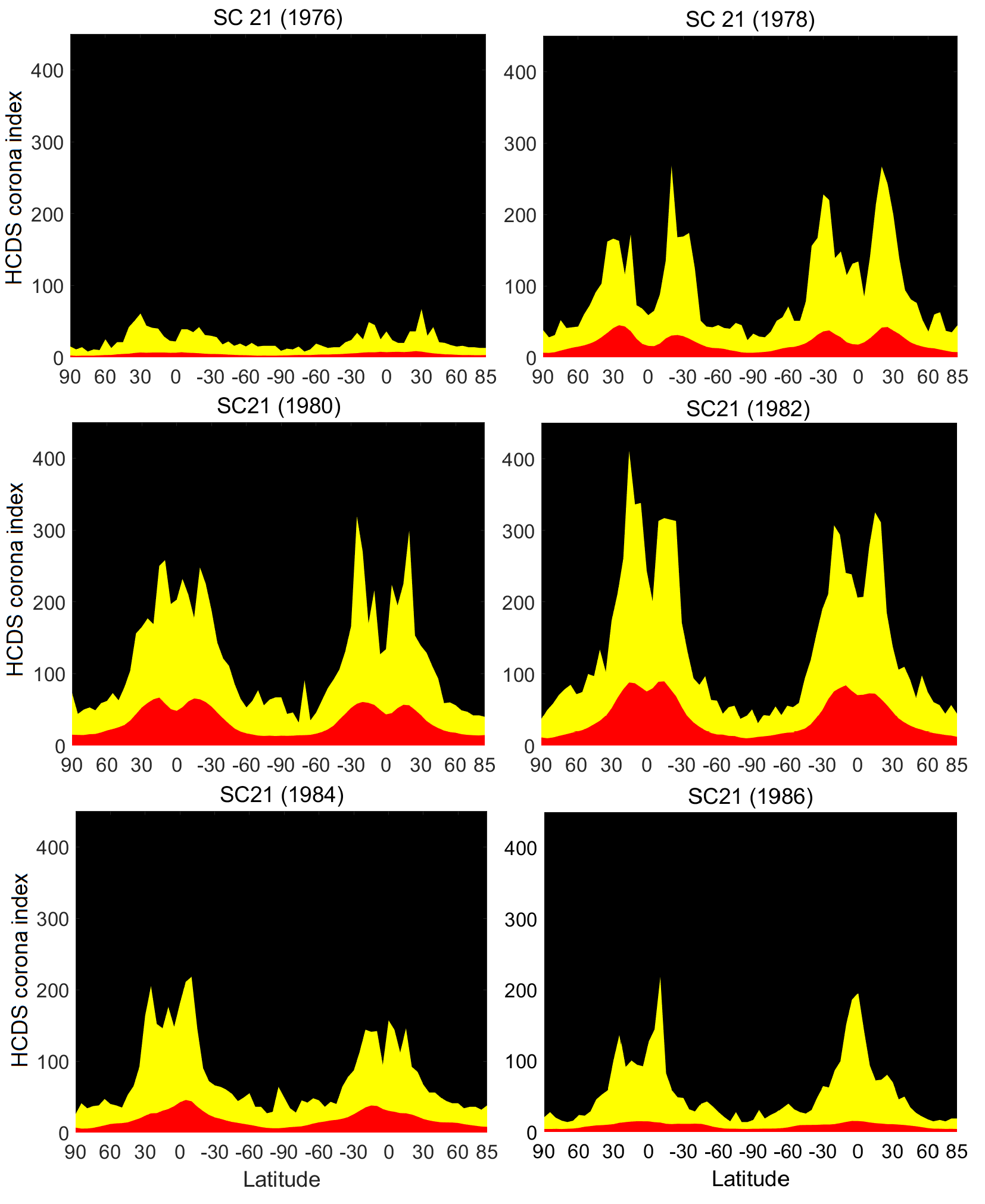}
		\caption{The maximum (yellow) and average (red) HCDS corona every second year for Solar Cycle 21.}
		\label{fig:SC_21_two_year_periods}
\end{figure}

Figure \ref{fig:SC_21_two_year_periods} shows the maximum (yellow) and average (red) HCDS corona every second year for Solar Cycle 21. The starting and ending minima for this cycle were June 1976 and September 1986, and the maximum December at the turn of the years 1979-80. The diagrams for 1976 and 1986 are calculated only for the partial years. As already seen from Fig. \ref{fig:Sunspots_and_corona_18_24}, the coronal maximum is not in 1980, but 1982, i.e. in the descending phase of the cycle. Another interesting feature is in the year 1984 diagrams. There are only two maxima in the average corona and both in the southern hemisphere. The maximum corona, however, exhibits peaks on both hemispheres.

\begin{figure}
	\centering
	\includegraphics[width=1.0\textwidth]{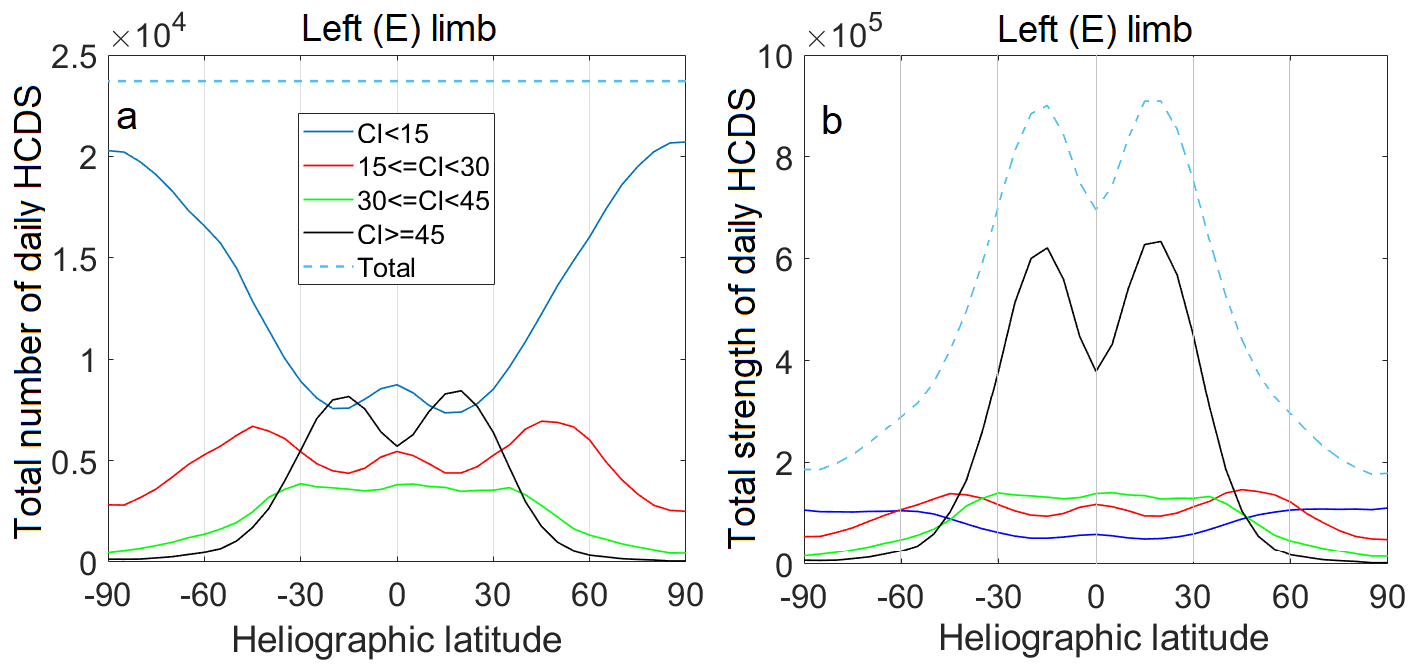}
		\caption{a) The amount of different category HCDS coronas (marked in the legend as CI) and b) the distributions of total strength of the categories as a function of heliographic latitude.}
	\label{fig:Total_number_of_corona}	
\end{figure}

Figures \ref{fig:Total_number_of_corona}a and \ref{fig:Total_number_of_corona}b show the number of different sizes of HCDS corona and total strength of these categories as a function of heliographic latitude on the left limb of the sun for Cycles 18\,--\,23 (we don't show here the right limb because distributions are understandably very similar). It is clear that the amount (Fig. \ref{fig:Total_number_of_corona}a) of the faintest corona ($<$15) is by far largest, except at the zones between 15\,--\,20 and -15\,--\,-20 of heliographic latitude. Furthermore, the amount of weak coronas increases towards the poles in both hemispheres. The latitudinal corona values between 15\,--\,30 are most abundant at 50 (-50) degrees of heliogaphic latitude and have also local maximum  at the solar equator. The next category (30\,--\,45) is almost constant between latitudes -30\,--\,30 and decreases towards the poles. The strongest coronas ($>=45$) are located mostly on the sunspot regions having maxima at the latitudes 15\,--\,20 and -15\,--\,-20. Fig. \ref{fig:Total_number_of_corona}b shows that, when calculating the total intensity of the categories, the strongest corona is dominating at the sunspot zone, i.e. between -45\,--\,45 degrees of latitude.

\begin{figure}
	\centering
	\includegraphics[width=1.0\textwidth]{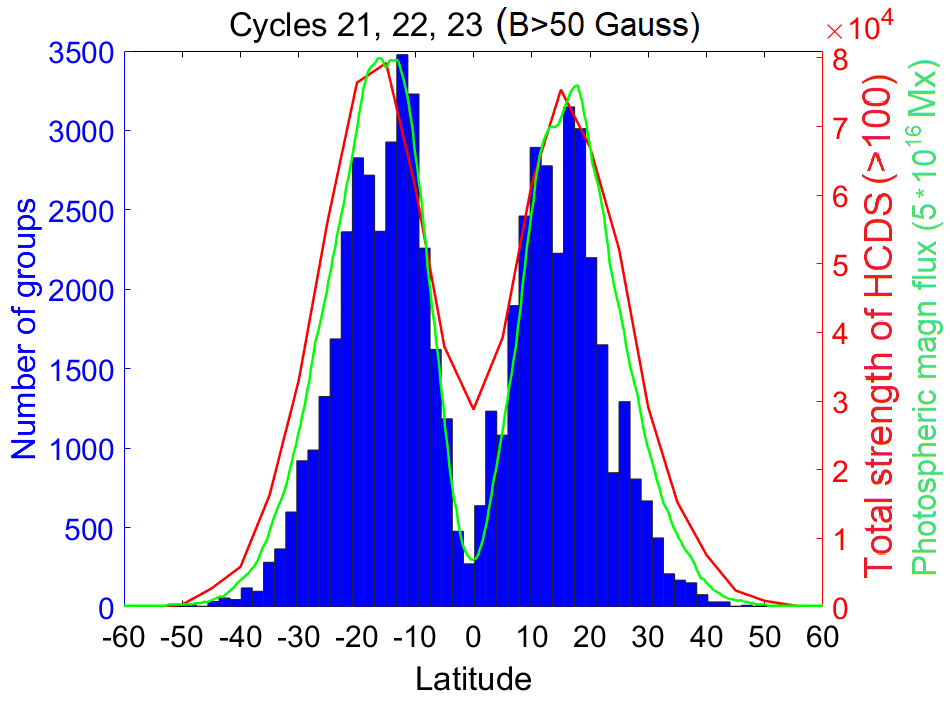}
		\caption{Number of HCDS corona events ($>$100, red) and photospheric magnetic flux (green) in Maxwells (Mx) for magnetic fields greater than 50 Gauss together with histogram of sunspot groups.}
		\label{fig:strongest_magnetic_field}
\end{figure}

\begin{figure}
	\centering
	\includegraphics[width=1.0\textwidth]{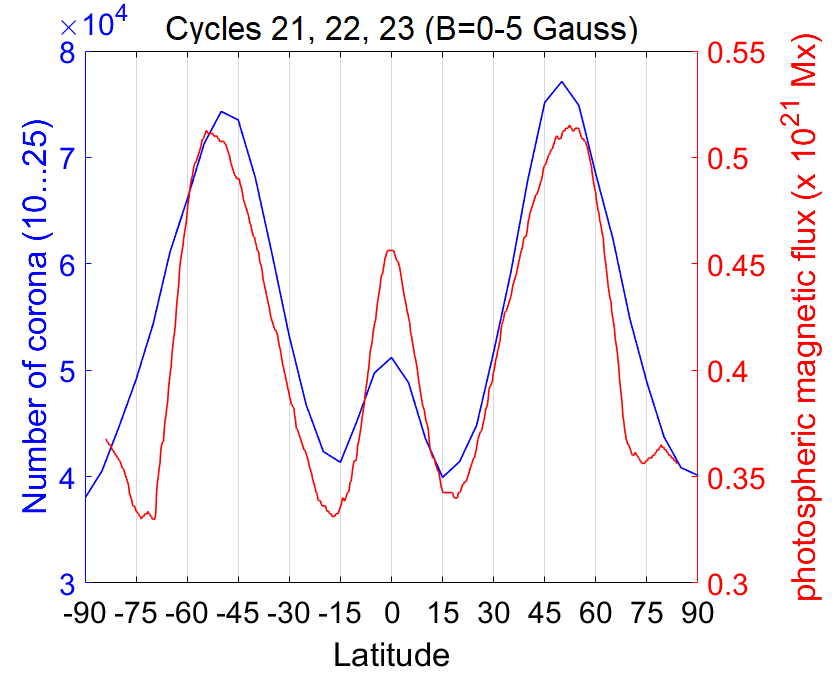}
		\caption{Number of HCDS corona events between 10\,--\,25 (blue) and photospheric magnetic flux (red) in Mx for magnetic fields between 0\,--\,5 Gauss.}
		\label{fig:faintest_magnetic_field}
\end{figure}

\cite{Vernova_2016} have studied the photospheric magnetic field distributions for Solar Cycles 21\,--\,23 in different categories of the magnetic field intensity (see also \cite{Rusin_2002}). Figure \ref{fig:strongest_magnetic_field} shows the distribution of the flux of the strongest photospheric magnetic fields (B$>$50 Gauss) as a green curve (y-axis also as green in the right side, in Maxwells, which is used in the paper of \cite{Vernova_2016}). This distribution is very similar to the distribution of the total strength of strongest HCDS corona events ($>$100, shown as red color) and the histogram of the amount of sunspot groups as a function of heliographic latitude (y-axis left). It should, however, be noted that the strongest magnetic field exist even at highest 15\% of the time in the photosphere (see more \citep{Vernova_2016}). It is interesting that the distribution of the flux of the faintest, i.e. the most abundant (35\,--\,80\% of the time) photospheric magnetic field (B=0\,--\,5 Gauss), has similar shape than the distribution of amount of the latitudinal coronas between 10\,--\,25. Figure \ref{fig:faintest_magnetic_field} shows the distribution of the amount of corona (blue) and photospheric magnetic flux (red) in Mx for these categories. The both curves have maxima at absolute values 50\,--\,55 and smaller maximum at the equator of the Sun.

\begin{figure}
	\centering
	\includegraphics[width=1.0\textwidth]{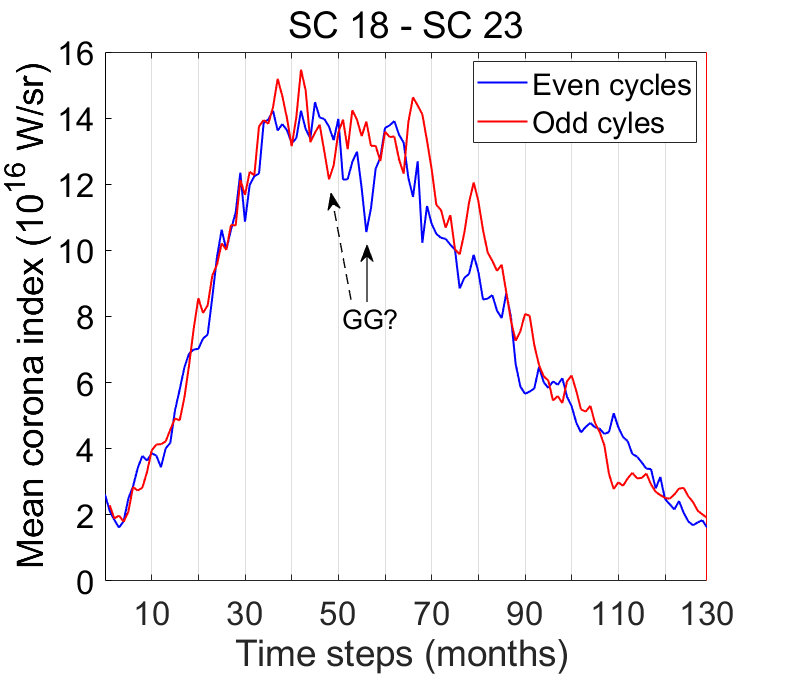}
		\caption{Average monthly CI for even and odd cycles of the period SC18\,--\,SC23}
		\label{fig:Monthly_corona}
\end{figure}

\section{Coronal Index of Solar Activity (CI)}

\subsection{The average shape CI cycle}

Figure \ref{fig:Monthly_corona} shows the monthly mean values of coronal index of solar activity (CI) for even and odd cycles between SC18\,--\,SC23, respectively. All the cycles are here resampled to have the same length of 130 time steps (months), which is about the average length of Solar Cycles 18\,--\,24. The shape of the monthly CI is very similar to the sunspot cycle with ascending phase about three years, quite flat top and five and a half year descending phase on the average. There is a good candidate for GG in the even cycles between 51\,--\,60 months, i.e. about 40\% from the start of the cycle. We calculated that two-sample T-test gives significance for the difference of the means in the interval 51\,--\,60 time steps compared to intervals 41\,--\,50 and 61\,--\,65 (note that the descending phase starts at about 65 time steps) with p=8.7$\times\!10^{-5}$. This means that significance is better than 99\% for even cycles.
The drop seen in odd cycles between 45\,--\,50, i.e. about 37\% from the preceding minimum, is less conspicuous (dashed arrow). However, the mean between 47\,--\,50 is significant at 95\% level with p=0.025 for four months compared to similar intervals before and after the gap.

\begin{figure}
	\centering
	\includegraphics[width=1.0\textwidth]{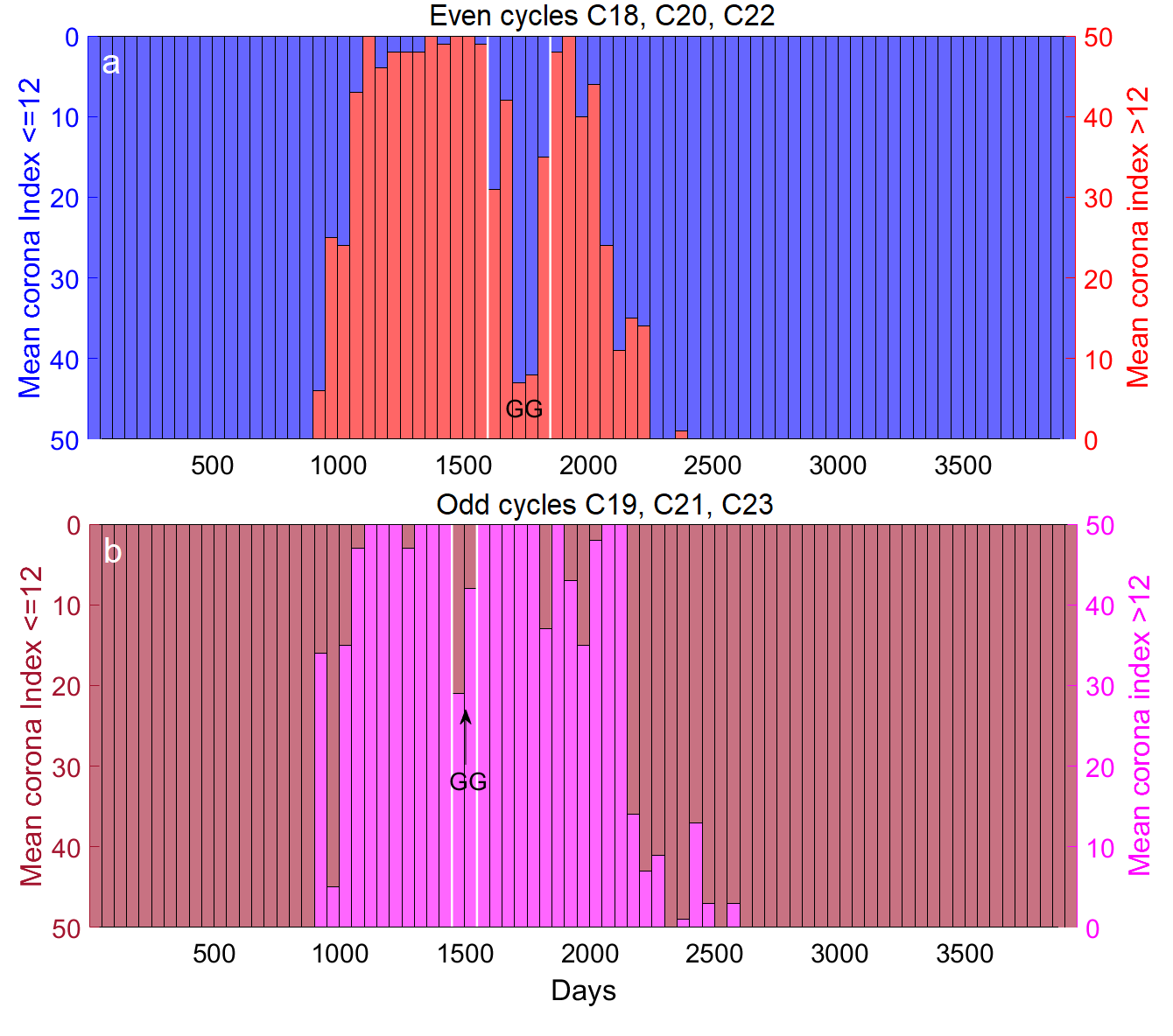}
		\caption{The amount of CI values $<=$12 or $>$12 as a histogram for a) even and b) odd cycles between SC18\,--\,SC23.}
		\label{fig:Histogram_daily_coronas}
\end{figure}

In order to study the relevance of the supposed GG in more detail, we use daily values of CI and calculate how many daily values are greater/less than half of the largest corona value (24.06) during the period of Cycles 18\,--\,23. Figures \ref{fig:Histogram_daily_coronas}a and \ref{fig:Histogram_daily_coronas}b show the histograms of daily CI values for even and odd cycles, respectively. We have resampled the daily values to 3950 days, which is about 130 months or 10.8 years, which is the average length of the Solar Cycles 18\,--\,23. In Fig. \ref{fig:Histogram_daily_coronas}a the CI values of even cycles under or equal 12 are shown as blue bars (y-axis as reversed on the left) and values over 12 as red bars (y-axis on the right). Each bar consists 50 days and because each day has one value the maximum is 50. In the middle of the maximum, i.e. with all values over 12, there is a deep gap with only a few large ($>$12) daily CI values (marked with white vertical lines in the figure). The deepest phase for this lasts over three months, but the decrease starts maybe another two months earlier. We have estimated using two-sample T-test that the mean value of the gap between the white lines ($\approx$ 8 months) has significantly different mean value (at level 99\%) with p = 0.0011 compared to equal period before or after the gap. We suppose that the period between the white lines is related to Gnevyshev gap (GG, marked in the \ref{fig:Histogram_daily_coronas}a). The gap is about 40\% from the preceding minimum of the average solar cycle. Note also the similarity of the average even cycle in Fig. \ref{fig:Monthly_corona} and the histogram of Fig. \ref{fig:Histogram_daily_coronas}a for even cycles. There exist first a smaller drop and then a deeper gap in both diagrams.

In Fig. \ref{fig:Histogram_daily_coronas}b for odd cycles the values under 12 are shown as light brown bars and values over 12 with magenta bars. Now the gap is somewhat earlier (37\% from the preceding minimum) and shorter ($\approx 3$ months). Although T-test gives again 95\% significance for the different mean, the result is somewhat suspicious, because of only two values in the gap. These both results are, however, very well in line with the earlier results for the GG \citep{Takalo_2020_a, Takalo_2020_c, Takalo_2021_a}.

\subsection{PCA of CI for Cycles 19\,--\,24}

\begin{figure}
	\centering
	\includegraphics[width=1.0\textwidth]{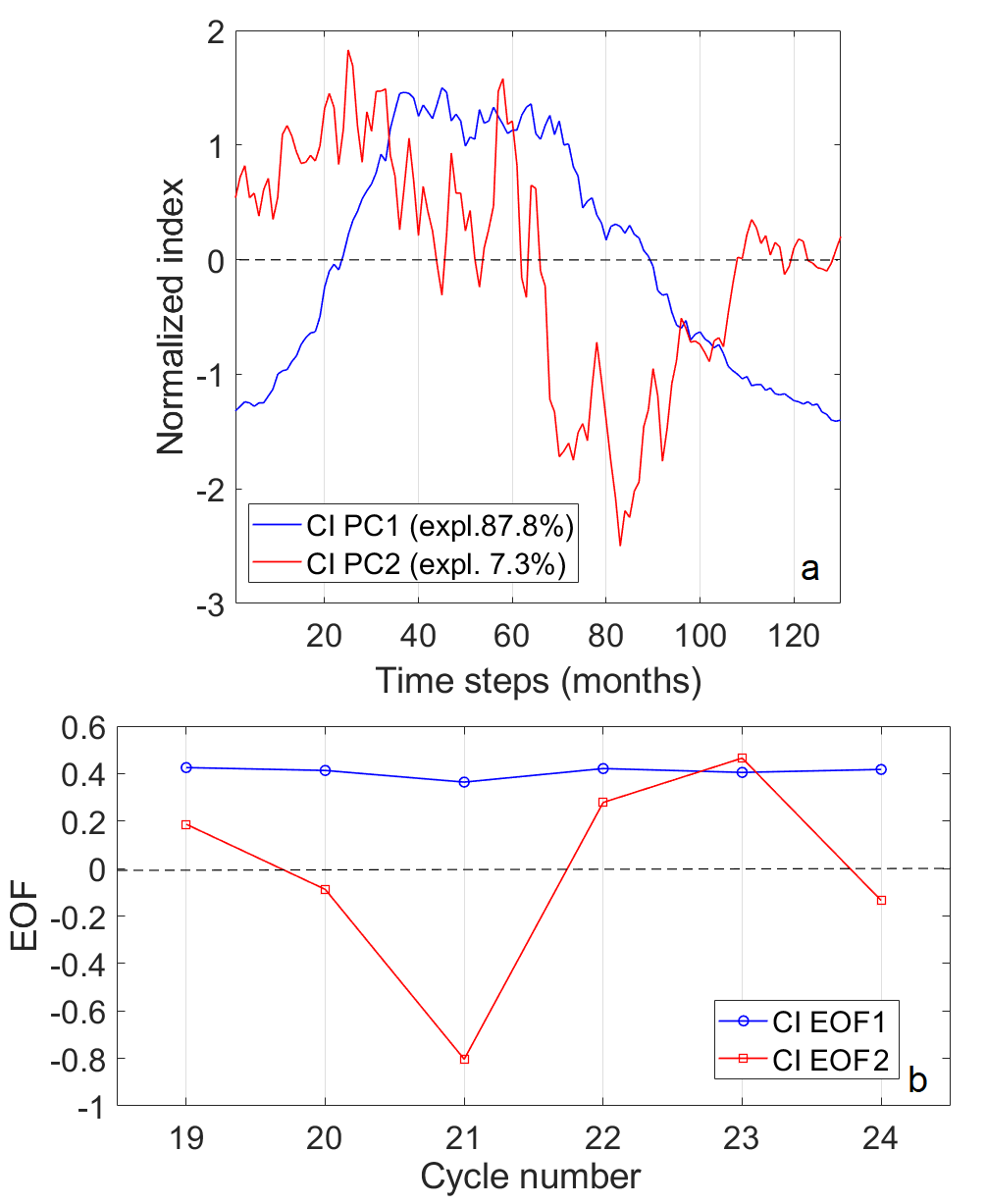}
		\caption{a) The PC1 and PC2 for coronal index of solar activity (CI). b) The corresponding EOF1 and EOF2 fot CI.}
		\label{fig:PCs_and_EOFs}
\end{figure}

\begin{figure}
	\centering
	\includegraphics[width=1.0\textwidth]{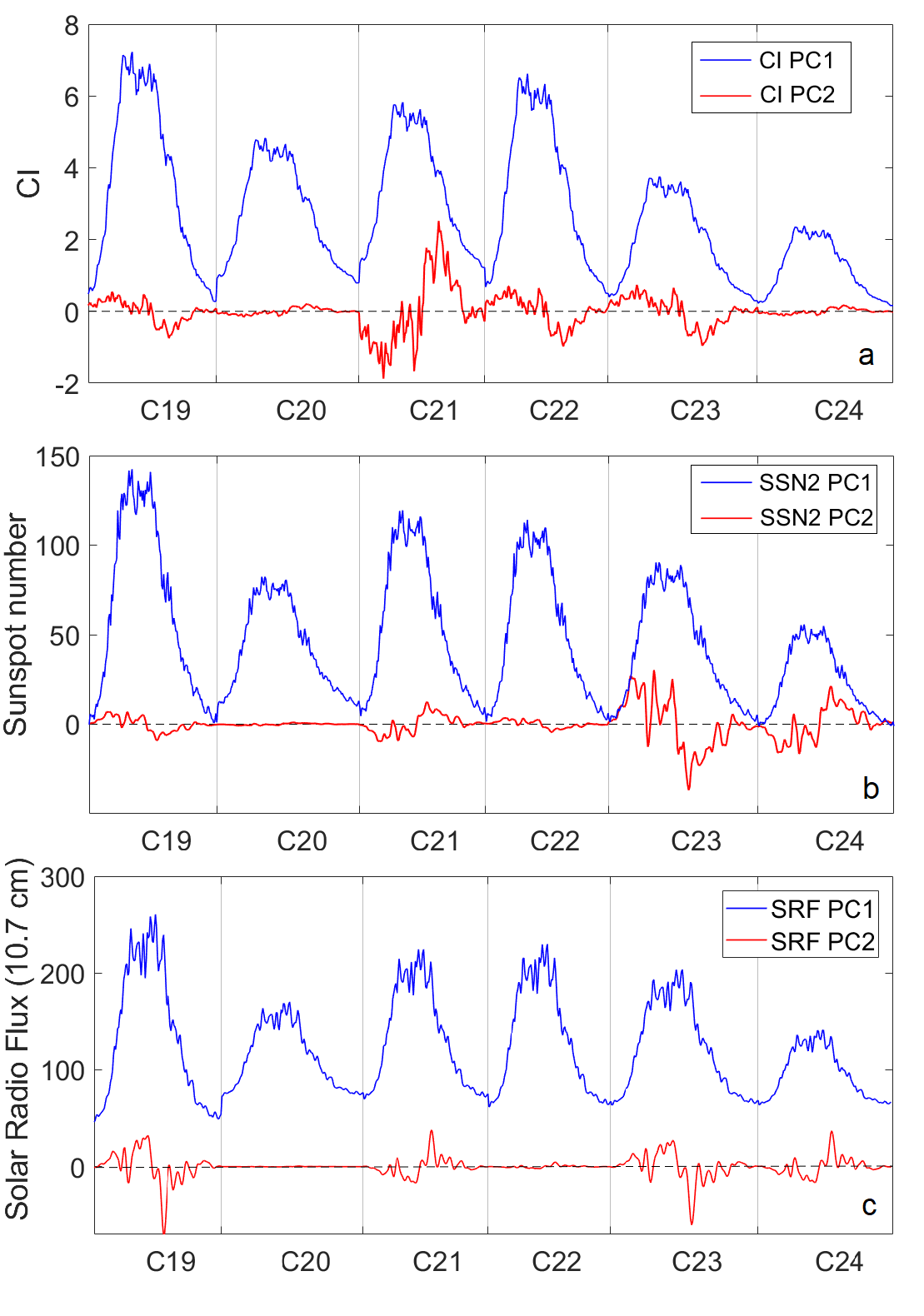}
		\caption{a) The PC1 and PC2 timeseries of CI for Cycles 19\,--\,24, b) The PC1 and PC2 timeseries of SSN for Cycles 19\,--\,24. c) The PC1 and PC2 timeseries of solar 10.7 cm radio flux for Cycles 19\,--\,24.}
		\label{fig:PC_time_series}
\end{figure}

\begin{figure}
	\centering
	\includegraphics[width=1.0\textwidth]{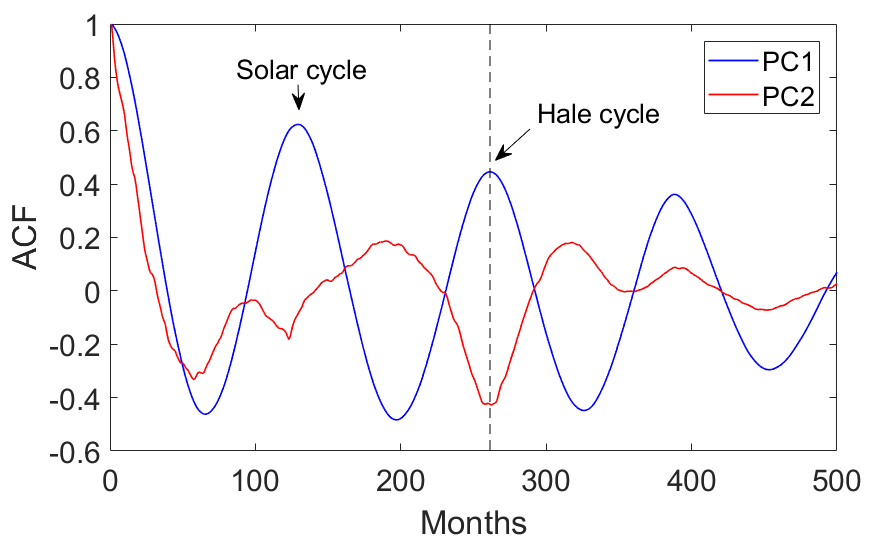}
		\caption{The autocorrelation functions of PC1 and PC2 timeseries of CI for Cycles 19\,--\,24. Note a deep minimum in the ACF of PC2 at Hale cycle period.}
		\label{fig:ACFs_of_PCs}
\end{figure}

We have carried out the principal component analysis by equalizing the CI Cycles to 130 time steps (months) to get the two main principal components (PC) shown in Fig. \ref{fig:PCs_and_EOFs}a. The first and second PC explain 87.8\,\% and 7.3\,\% of the total variation of the CI  period C19\,--\,C24, i.e together 95.1\,\% of the variation. The rest four PCs account only for 4.9\% of the variation and are usually just explaining some special feature of an individual cycle. The corresponding EOFs are shown in Fig. \ref{fig:PCs_and_EOFs}b. According to theory of PCA, PC1 should show the average shape of the cycles for the studied interval. The PC2 acts to correct the shape of the cycle when the corresponding cycle differs from the average cycle shape. The main effect of the PC2 is to reduce (positive scaling for PC2) or enhance (negative scaling for PC2) the activity level of the declining phase with respect to the ascending phase of the cycle \citep{Takalo_2018}. Positive (negative) scaling of PC2 means that the it has positive (negative) phase in the first half of the cycle and negative (positive) phase in the second half of the cycle. Now looking at the EOFs of Fig. \ref{fig:PCs_and_EOFs}b it shows that the CI Cycle 21 differs most of the other cycles such that it has least weight to the PC1 (its EOF1 is smallest), but by far highest (but negative) weight in its EOF2. We then returned each cycle to its original length, and back to its original amplitude by multiplying both PCs with the standard deviation of the original cycle and and adding the mean value of the original cycle to PC1. Then we concatenated the cycles to their original order and obtained the full PC1 and PC2 series of Cycle 19\,--\,24. These two time series are shown in Fig. \ref{fig:PC_time_series}a. Note that indeed CI Cycle 21 has highest amplitude in PC2 timeseries with negative phase in the first half and positive phase in the latter half of the cycle. The peak-to-peak variation of PC2 for the Cycle 21 is more than half of the height of the PC1 for that Cycle. This confirms the characteristic of the CI Cycle 21 of Fig. \ref{fig:Sunspots_and_corona_18_24} and \ref{fig:HCDS_lines}. The corona (both HCDS and CI) is strongly enhanced in the descending phase of the Cycle 21. The Cycles 20 and 24 have only slightly negative scaling. The other cycles, i.e. C19, C22 and C23, have positive scaling meaning that the first half, i.e. the ascending phase of these cycles has stronger corona than the average cycle.
Figures \ref{fig:PC_time_series}b and c show similar PC analyses for Sunspot number (SSN) and Solar 10.7 cm Radio Flux (SRF) data for Cycles 19\,--\,24, respectively. (The reason why we study only Cycles 19\,--\,24 is that SRF data does not consist the whole Cycle 18.)  It is evident that these two data have quite similar PC2 time series with the same phases as CI index, although the strength of the peak to peak variation changes somewhat. Note, that the PC2 of Cycle 20 for all three data is almost a zero line, meaning that the PC1 is kind of a 'model' cycle of these data for the interval C19\,--\,C24. The variation PC2 for CI Cycle 24 is also quite minimal, although it fluctuates quite a lot with negative scaling for SSN and SRF. Note also that the variation of the PC2 of the Cycle 21 is much weaker for SSN and SRF than for CI. There is kind of a 22-year (Hale cycle) anticorrelation, such that PC2 of CI has positive scaling for cycle 19 (more weight on the first half of the cycle), negative scaling for cycle 21 (more weight on the latter half) and again positive scaling for cycle 23 (more weight on the first half). Similarly, C20, C22 and C24 have negative, positive and negative scaling, respectively.
Figure \ref{fig:ACFs_of_PCs} shows the autocorrelation functions (ACFs) of the PC1 and PC2 of CI for Cycles 19\,--\,24. The ACF of PC1 shows, as expected, maxima at the solar cycle variation in CI. The PC2 is more complicated and shows maxima near 1.5 and 2.5 solar cycle periods. Interestingly, there is a deep minimum in ACF of PC2 at Hale cycle period (262 months or 21.8 years). This confirms the aforementioned anticorrelation at Hale cycle in the CI. \newline

\section{Conclusions}

We found that the overall values of the homogeneous corona data set (HCDS) corona events are slightly stronger for northern hemisphere than the southern hemisphere during Solar Cycles 18\,--\,23. It is noticeable that the Cycles 19 and 22 are most asymmetrical such that northern hemisphere is dominating for Cycle 19 and southern hemisphere is dominating for Cycle 22. These are, however, also the cycles, which are the most symmetric around the cycle maximum for strong HCDS corona events. The corona values for Solar Cycles 23 and 24 are understandably by far smaller than for the other cycles.

The maximum HCDS corona maximum appears first around 1000 days, i.e. about a quarter after the start of the cycles near 40\,--\,50 degrees of northern latitude and somewhat later in the southern hemisphere at corresponding latitudes. These maxima migrate towards the poles such that they appear at the pole around 1200 days and 1350 days after the start of the cycle for northern and southern hemisphere, respectively. The equatorward drifts reach the low latitudes at about half of the cycle, and the maximum is clearly double-peaked for the average even cycle.

The amount of weak HCDS corona increases towards the poles in both hemispheres. The corona values
between 15\,--\,30 are most abundant at 50 (-50) degrees of heliographic latitude and have also local maximum at the solar equator. The next category (30\,--\,45) is almost constant between latitudes -30\,--\,30 and decreases towards the poles. The strong HCDS corona ($\geq$45) are located mostly on the sunspot regions having
maxima at the absolute value of latitudes 15\,--\,20 in both hemispheres. When calculating
the total intensity of the categories, the strong corona events are dominating at the
sunspot zone, i.e. between -45\,--\,45 degrees of latitude.

The distribution of the flux of the strongest photospheric magnetic fields (B$>$50 Gauss)
is very similar to the distribution of the total strength of strongest corona events
($>$100) and the histogram of the amount of sunspot groups. It should, however, be noted
that the strongest magnetic field exist even at highest 15\,\% of the time in the
photosphere.

We found Gnevyshev gap in corona index of solar activity (CI) for the even cycles between 51\,--\,60 months, i.e.
about 40\,\% from the start of the cycle. We calculated using two-sample T-test at least
99\,\% significance for the difference of the means in the interval 51\,--\,60 time steps
compared to intervals 41\,--\,50 and 61\,--\,65 with p=8.7$\times$ 10$^{-5}$. We confirmed the existence GG 
also for daily CI data.

For the odd cycles the GG exists little earlier and is shorter than for the even cycles. 
Although the significance of the gap is at 95\,\% level for the odd cycles, its existence is
vague compared to the even cycles.

We have carried out the principal component analysis (PCA) of the CI for Cycles 19\,--\,24. The principal components PC1 and PC2 account for 87.7\,\% and 7.3\,\% of the total variation of the data. The PCA confirms that Cycle 21 is most different from other cycles such that its PC2 has strong negative scaling, i.e. negative phase in the first half and positive phase in the latter half of the cycle. We compared the PC1 and PC2 of the CI to the PC1 and PC2 of sunspot numbers and solar 10.7 cm radio flux data. We found that their PC2 resembles that of the PC2 for CI such that they are in the same phase for all cycles, which differ from zero line.

\begin{acknowledgments}
The dates of cycle minima were obtained from from the National Geophysical Data Center, Boulder, Colorado, USA (ftp.ngdc.noaa.gov). The newly reconstructed corona indices were fetched from www.ngdc.noaa.gov/\newline stp/solar/corona.html. The corona index for Solar Cycle 24 was fetched from \newline www.suh.sk/obs/vysl/MCI.htm. The SSN data has been downloaded from \newline www.sidc.be/silso/datafiles and the Solar radio flux data from  lasp.colorado.edu \newline /lisird/data/penticton\_radio\_flux/. The author is grateful for the reviewer of the constructive advice.
\end{acknowledgments}

\flushleft
\textbf{Disclosure of Potential Conflicts of Interest} \newline
\footnotesize {The author declares that there are no conflicts of interest.}

%
%
\bibliographystyle{spr-mp-sola}
\bibliography{references_JT_SolPhys}  
%
%
%
%


\end{document}